# Correlation between spin state and activity for hydrogen evolution

Tao Zhang[1], Lei Li[1], Tao Huang[1], Hui Wan[2,1], Wu-Yu Chen[1], Zi-Xuan Yang[1], Gui-Fang Huang[1#],

Wangyu Hu[3], Wei-Qing Huang[1*]

[1]Department of Applied Physics, School of Physics and Electronics, Hunan University, Changsha 410082, China

[2]School of Materials and Environmental Engineering, Changsha University, Changsha, 410082, China.

[3]College of Materials Science and Engineering, Hunan University, Changsha, 410082, China.

**Abstract**: Spin plays a key role in physical and chemical reactions, such as oxygen evolution and hydrogen evolution reactions (OER/HER); but the spin-activity correlation has remained unclear. Based on a transition metal (TM)-doped $PtN_2$ monolayer model with a well-defined spin center as adsorption site, we here reveal that only active spin state can enhance the strength of hydrogen adsorption, while inert spin state offers very little influence. Specifically, the unpaired electron along the out-of-plane direction such as in $d_{z^2}$ orbital, acting as an active spin state, will strongly hybridize with hydrogen, resulting in enhanced hydrogen binding energy because $d_{z^2}$ orbital is just enough to accommodate two electrons to form a bonding orbital. While the in-plane unpaired electron such as in $d_{x^2-y^2}$ orbital, plays a negligible role in adsorbing hydrogen atom. This is verified by a series of single atom catalysts comprising of $PtN_2$ monolayer by replacing Pt atom with a TM (Fe, Co, Ni, Ru, Rh, Pd, Os, or Ir) atom, or subsequent adsorbing a Cl atom. One of the most promising materials is Pd@$PtN_2$-Cl that offers superior HER activity, even better than pure Pt. This work uncovers the nature of spin-activity correlation, thus paving the way for the design of high-performance catalysts through spin-engineering.

---

[*]. Corresponding author. *E-mail address:* wqhuang@hnu.edu.cn

[#]. Corresponding author. *E-mail address:* gfhuang@hnu.edu.cn

## I. Introduction

Like charge [1–6], spin is an intrinsic property of electrons and plays a key role in physics and chemistry with a broad range of applications including quantum devices, spintronics, and materials science [7–17]. Recently, spin has emerged as a pivotal determinant influencing chemical reactions, such as oxygen evolution and hydrogen evolution reactions (OER/HER) [18–27]. For instance, spin-polarized electrons in ferromagnetic oxides (such as, $NiZnFeO_x$, $Co_{3-x}Fe_xO_4$) promote the production of parallel spin-arranged oxygen through quantum spin exchange interaction (QSEI), greatly improving the OER kinetics in alkaline solutions [23–25]. Moreover, the electronic spin polarization can also increase the overlap-integral between the catalysts and the precursor/intermediate/product species to enhance charge transfer, therefore modifying the binding energy [26,27]. A particular case is carbon-based magnetic catalytic nanocages (N, F, B doped carbon nanofibers with cobalt nanodots), in which the possible electron transition of cobalt atoms from low spin to high spin could be occurred under an applied magnetic field, thus increasing the spin polarization and producing more unpaired electrons at the $3d$-orbit of Co [27]. This can effectively enhance the adsorption of oxygen intermediates and electron transfer during the reaction. However, the materials synthesized by most experimental methods are not an ideal model system to explore the spin-activity nature due to their inherent structural complexity, the ambiguous local coordination environment and a lack of the ability to control the spin states of active sites [28].

Fortunately, the single-atom catalysts (SACs) have a well-defined active site and its electronic structure is relatively easy to regulate [29–33]. In this respect, the SACs may be an ideal model system to study the relationship between the spin state of the active site and the catalytic activity. Particularly, the magnetic SACs (MSACs) with magnetic active centers are extremely important for revealing the microscopic mechanism of the interaction between adsorbed intermediates and the spin state of reaction sites in catalytic reactions [34–36]. Moreover, external field or energy is no longer needed to create unpaired electrons because the MSACs have the well-defined, intrinsic magnetic centers (sites). This bears out that spin-engineering, as an effective strategy to improve the performance of catalysts, is particularly interesting both from fundamental and practical points of view. Unveiling the nature of spin-activity correlation is therefore an essential prerequisite to design spin catalysts with high performance.

Herein, we explore the nature of spin-activity correlation and find that spin state can be divided into two categories: active spin state and inert spin state, the former of them could enhance the hydrogen binding energy, while the latter has little effect. Starting from a planar quadrilateral $PtN_2$ monolayer with no initial magnetic moment, transition metal (TM = Fe, Co, Ni, Ru, Rh, Pd, Os, or Ir) atom is doped into the $PtN_2$ monolayer by replacing a Pt atom, or further adsorbing a Cl atom, to induce an unpaired electron, obtaining a series of $PtN_2$-based MSACs with definite magnetic active sites. We find that the unpaired electron along the out-of-plane direction such as in $d_{z^2}$ orbital of TM atom is an active spin state, which will bond with H atom, thus strengthening the binding energy; while the in-plane unpaired electron such as in $d_{x^2-y^2}$ orbital of TM atoms, has very weak influence on the adsorbed hydrogen atom. One of the most promising catalysts is Pd@$PtN_2$-Cl that shows superior HER catalytic activity, even better than the benchmark Pt [37]. Our work reveals the correlation between spin state and HER activity.

## □. Computational Methods

Our spin-polarized DFT calculations [38] were performed by using Vienna ab initio simulation package (VASP) [39,40]. The projector augmented wave potentials are utilized to deal with the interaction between core electrons [41]. The electron-ion and the exchange-correlation interactions are described by the Perdew-Burke-Ernzerhof (PBE) [42]. The van der Waals interaction is described by using the zero-damped DFT+D3 method [43,44]. And the energy cutoff of the plane-wave is set to be 550 eV. All considered structures are fully relaxed under the convergence criteria of $10^{-5}$ eV and 0.01 eV/Å of total energy and atomic force, respectively. The unit cell structures of monolayer $PtN_2$ were relaxed with a $\Gamma$-centered Monkhorst−Pack k-point mesh [45] of 9×7×1 in the first Brillouin zone. To avoid interactions between adjacent periodic images of the monolayers, we inserted a vacuum region of 20 Å along the plane normal direction. The thermodynamic stability of $PtN_2$ were also examined by ab initio molecular dynamic simulations performed at 300 K [46]. In addition, the hybrid Heyd-Scuseria-Ernzerhof (HSE) method is used to correct the band structure of $PtN_2$ unit cell [47,48].

At standard conditions ($T$ = 298.15 K, $P$ = 1.01×105 Pa) and potential of U = 0, the overall HER process can be written by

$$H^+ \ (aq) \ + e^- \rightarrow \frac{1}{2}H_2(g) \qquad \Delta G = 0\text{eV} \qquad (1)$$

This reaction contains the initial reactants $H^+$ and $e^-$, the intermediate adsorbed $H^*$, and the final product $H_2$. The total energies of $H^+(aq) + e^-$ and $\frac{1}{2}H_2(g)$ are equal. Therefore, according to the Sabatier principle, the Gibbs free energy of atomic H adsorption $(\Delta G_H)$ is a good descriptor to evaluate HER activity of catalysts and is given by [49]

$$\Delta G_H = \Delta E_{H^*} + \Delta E_{ZPE} - T\Delta S \qquad (2)$$

where $\Delta E_{ZPE}$ is the correction of zero-point energy and $T\Delta S$ is the entropy difference between adsorbed $H^*$ and $H_2$ in gas phase at 298.15 K. $\Delta E_{H^*}$ is the H adsorption energy, which can be calculated by,

$$E_{H*} = E_{slab+H^*} - E_{slab} - \frac{1}{2}E_{H_2} \qquad (3)$$

where $E_{slab+H^*}$, $E_{slab}$ and $E_{H_2}$ are total energies of monolayer slab with one H atom adsorbed on the surface, slab, and $H_2$ molecule, respectively.

The formation energies $(E_f)$ of the TM@PtN$_2$ systems were calculated for comparing their relative stability by the equation:

$$E_f = E_{total} - E_{defect} - E_{TM} \qquad (4)$$

where $E_{total}$ is the total energy of TM@PtN$_2$, $E_{defect}$ is the total energy of PtN$_2$ with one Pt vacancy and $E_{TM}$ is the energy of one TM atom.

The adsorption energies $(E_{ad})$ of Cl atom in TM@PtN$_2$-Cl systems were calculated by the equation of

$$E_{ad} = E_{slab+Cl^*} - E_{slab} - E_{Cl} \qquad (5)$$

where $E_{slab+Cl^*}$, $E_{slab}$ and $E_{Cl}$ are total energies of TM@PtN$_2$-Cl, TM@PtN$_2$ and Cl atom, respectively.

The transition metal−H (TM−H) interaction strength is described by crystal orbital Hamilton population (COHP) [50] which can be calculated using the implementation in the Local Orbital Basis Suite Toward Electronic-Structure (LOBSTER) package [51].

## Ⅲ. Results and Discussions

As a starting point, we construct a planar quadrilateral PtN$_2$ monolayer and the optimized structure of PtN$_2$ with 3×3×1 supercell is shown in Fig. 1(a). It can be found that after optimization the PtN$_2$ monolayer, like graphene, is a planar structure, in which all atoms are in the same plane, each Pt atom is surrounded by four coordinated N atoms, and its lattice constants are a = 2.02 Å and b = 1.24 Å. The electronic band structure for the PtN$_2$ unit cell is shown in Fig. 1(b). We note that the calculated energy band is continuous near the Fermi level, indicating that the PtN$_2$ monolayer have fairly good electrical conductivity. The phonon spectrum of the PtN$_2$ monolayer shows that it is dynamically stable in view of the absence of imaginary frequencies within the first Brillouin zone (Fig. S1(a)) [52]. Meanwhile, to assess the thermal stability of PtN$_2$, AIMD simulations with time of 5 $ps$ were employed. The results display that PtN$_2$ exhibits no obvious distortion in geometrical structure at room temperature (300 K), and the corresponding results of total energy oscillate around a constant value, demonstrating the thermal stability of PtN$_2$ (Fig. S1(b)).

To determine optimal H adsorption site of the PtN$_2$ monolayer surface, five potential adsorption sites are estimated, including the top sites of N atom (T$_N$) and Pt atom (T$_{Pt}$), the hollow site of Pt-N six-member ring (H$_{Pt-N}$), the bridge sites of N-N bond (B$_{N-N}$) and Pt-N bond (B$_{Pt-N}$) (Fig. S2). When H atom is adsorbed on each of these positions, only T$_N$, T$_{Pt}$ and H$_{Pt-N}$ sites can remain the adsorption behavior after structural optimization. The Gibbs free energy ($\Delta G_H$) of H atom was calculated to examine the activity of HER, as shown in Fig. 1(c). Among all these sites, the T$_{Pt}$ site exhibits the relatively good performance; but even so, its $\Delta G_H$ value (0.89 eV) is still much higher than 0.2 eV, suggesting that the adsorption of H atom is too weak on the PtN$_2$ monolayer surface.

The PtN$_2$ monolayer is a non-magnetic structure because the arrangement of electrons in the spin channel is symmetric and there is no unpaired electron. Recent studies have shown that spin plays a key role in physical and chemical reactions [18–27]. For instance, the applied magnetic field promotes to produce more unpaired electrons at the Co-3$d$ orbitals in carbon-based magnetic catalytic nanocages, effectively enhancing the adsorption of oxygen intermediates [24]. Therefore, it is reasonable to speculate that the unpaired electron would also promote the adsorption of H atom. To verify this conjecture, the electron configuration of the Pt atom in PtN$_2$ is first studied. Each PtN$_2$ unit cell contains one Pt atom and two N atoms, and the adjacent N atoms between two unit cells share two electron pairs. When the Pt atom provides an electron to each N atom, the N atom will reach the eight-electron stable structure. In this case, each Pt atom ends up with eight valence electrons. According to the crystal field theory, for the center ion in the quadrilateral coordination field, its $d_{xz}$ and $d_{yz}$ orbitals are degenerate and have the lowest energy, and then the energy of $d_{x^2-y^2}$, $d_{z^2}$, and $d_{xy}$ orbitals are arranged in order from low to high. As a result, the Pt atom has $d^8$ electron configurations (the left panel of Fig. 1(d)). Since Ni, Pd and Pt have the same number of valence electrons, the electron configurations of the Ni and Pd atoms are therefore the same as that of the Pt atom when the Pt atom is substituted by Ni or Pd atom. To induce an unpaired electron in the central atom, we design two schemes, as shown in the middle panel of Fig. 1(d). The first scheme is doping, that is replacing the Ni, Pd, or Pt with an atom in the previous family of them (such as Co, Rh, Ir), so that the central atom in quadrilateral coordination field will have seven electrons, leaving the $d_{z^2}$ orbital with an unpaired electron. Considering the Cl atom is highly electronegative and has seven valence electrons, the second scheme is to adsorb a Cl atom below the central atom in the Ni@PtN$_2$, Pd@PtN$_2$ and PtN$_2$ systems. In this way, the less electronegative central atom acts as an electron donor, giving one electron to the more electronegative Cl atom. In this case, it also leaves the central atom with seven electrons, inducing an unpaired electron in the $d_{z^2}$ orbital, as shown in the right panel of Fig. 1(d).

The doping scheme is firstly investigated. Fig. 2(a) shows that the planar structure does not change significantly after doping, and all atoms are still in a same plane. To verify the stability of TM@PtN$_2$ systems, the formation energies $\left(E_f\right)$ are calculated. The $E_f$ of TM@PtN$_2$ varies with the doping atom, and it ranges from -2.09 eV to -0.52 eV (Fig. S3), indicating that these systems are stable.

As expected, the calculated results of magnetic moments show that when the Ni, Pd, or Pt atom is replaced by the Co, Rh, or Ir atom, the systems are induced to be magnetic (the magnetic moment of Co@PtN$_2$, Rh@PtN$_2$, and Ir@PtN$_2$ systems are $0.98\mu_B$, $0.61\mu_B$, and $0.35\mu_B$, respectively), as shown in Fig. 2(b1). This indicates that the central atom has an unpaired electron. Based on this transition, we can investigate the relationship between the spin and HER activity of these systems. The $\Delta G_H$ is used as a descriptor for activity, and the calculated $\Delta G_H$ are shown in Fig. 2(b2). Obviously, compared with non-magnetic systems, the magnetic systems in which the $d_{z^2}$ orbital of central atom contains an unpaired electron exhibit much lower $\Delta G_H$, indicating that the adsorption of H atom is enhanced. Taking non-magnetic Ni@PtN$_2$ and magnetic Co@PtN$_2$ as examples, the partial density of states (PDOS) of the central atom are calculated and shown in Fig. 2 (c). We find that the spin state contributed by Co-3$d_{z^2}$ orbital appears near the Fermi level (the green part of the figure), but such spin state is not found in the PDOS of Ni atom. This verifies that the unpaired electron exists in the Co-3$d_{z^2}$ orbital. There similar results for other systems (such as Pd@PtN$_2$, Rh@PtN$_2$, PtN$_2$ and Ir@PtN$_2$) are shown in Fig. S4. In addition, we also find that for the magnetic systems (such as Co@PtN$_2$), the H-1$s$ electron mainly interacts with the Co-3$d_{z^2}$ electron, and the magnetic moment of the system becomes zero after the adsorption of H atom (Fig. 2(d)). This is also true for other magnetic systems (Fig. S5). These results demonstrate that the unpaired electron in the $d_{z^2}$ orbital of TM atom is an active spin state that can bond with H atom to increase the TM-H interaction.

We next study the effect of the addition of a Cl atom on the H atom adsorption. For comparison, a Cl atom is adsorbed to not only the non-magnetic systems (i.e., Ni@PtN$_2$, Pd@PtN$_2$ and PtN$_2$), but also the magnetic systems (i.e., Co@PtN$_2$, Rh@PtN$_2$ and Ir@PtN$_2$). The top view of TM@PtN$_2$-Cl is similar to TM@PtN$_2$, and its side view is displayed in Fig. 3(a). The axially bonded Cl ligand pulls the central atom into a valley site, resulting in the central atom deviation from PtN$_2$ plane. The elongated TM-N bonds drop the energy level of $d_{x^2-y^2}$ and $d_{xy}$, and the replacement of central atom results in the elevated energy levels of $d_{xz}$, $d_{yz}$, and $d_{z^2}$. Fortunately, the movement of energy level is not enough to change the relative positions of the sub-orbitals. The calculated results of the charge density difference and Bader charge clearly show that electron transfers from the central atom to the Cl atom, which confirms our initial assumption, as shown in Fig. 3(b) and Fig. S7. As a result, Ni@PtN$_2$-Cl, Pd@PtN$_2$-Cl and PtN$_2$-Cl systems have an unpaired electron in $d_{z^2}$ orbital of the central atom (the right panel of Fig. 1(d)); whereas Co@PtN$_2$-Cl, Rh@PtN$_2$-Cl and Ir@PtN$_2$-Cl systems do not have electrons in $d_{z^2}$ orbital (Fig. S6). Similar to the case of doping systems, the $E_{ad}$ of Cl atom in TM@PtN$_2$-Cl is calculated to verify the stability of the systems. As shown in Fig. 3(c), the value of E$_{ad}$ ranges from -4.88 eV to -3.71 eV, demonstrating that the adsorption of Cl atom is stable in these systems. Furthermore, we are surprised to find that the adsorption of Cl atom in magnetic systems (Co@PtN$_2$, Rh@PtN$_2$ and Ir@PtN$_2$) are stronger than in non-magnetic systems (Co@PtN$_2$, Rh@PtN$_2$ and Ir@PtN$_2$), which is consistent with the case of H atom adsorption.

The calculated magnetic moment and $\Delta G_H$ of TM@PtN$_2$-Cl are shown in Fig. 3(d1) and (d2). The relationship between spin and activity is also same as that mentioned in doping systems. For those magnetic TM@PtN$_2$-Cl systems, the $\Delta G_H$ of Ni@PtN$_2$-Cl, Pd@PtN$_2$-Cl, and PtN$_2$-Cl is reduced by 0.76 eV, 1.09 eV, and 1.39 eV compared to the non-adsorbed Cl atom systems, indicating that the adsorption of H atom is enhanced by Cl atom. Especially, Pd@PtN$_2$-Cl endows the best HER activity with an optimal $\Delta G_H$ of 0.08 eV, a value better than pure Pt (0.09 eV). In contrast, for non-magnetic TM@PtN$_2$-Cl systems, the $\Delta G_H$ of Co@PtN$_2$-Cl, Rh@PtN$_2$-Cl and Ir@PtN$_2$-Cl is increased by 0.12 eV, 0.41 eV, and 0.20 eV compared to the non-adsorbed Cl atom systems, suggesting that the adsorption of H atom is weakened by Cl atom. Similarly, the PDOS calculation results of central atoms show that there are "active spin states" near the Fermi level contributed by $d_{z^2}$ orbitals for the magnetic TM@PtN$_2$-Cl systems, but not for the non-magnetic TM@PtN$_2$-Cl systems, as displayed in Fig. 3(e) and Fig. S8. At the same time, the adsorbed H atom mainly interacts with TM-$d_{z^2}$ orbital, and the magnetic moment of the magnetic TM@PtN$_2$-Cl become zero after the adsorption of H atom (Fig. 3(f) , Fig. S9), in agreement with the results in magnetic TM@PtN$_2$.

Interestingly, the situation is quite different when a Cl atom is adsorbed to the central atom of Fe@PtN$_2$, Ru@PtN$_2$, or Os@PtN$_2$. As depicted in Fig. 4(a), the influence of Cl atom on the magnetic moment of Fe@PtN$_2$ and Ru@PtN$_2$ is different from that of Os@PtN$_2$, due to the splitting energy difference of the central atomic orbitals. The electron configurations of the central atom in the Fe@PtN$_2$, Fe@PtN$_2$-Cl, Ru@PtN$_2$, and Ru@PtN$_2$-Cl systems are different (Fig. S10), which can be confirmed by their PDOS (Fig. S11). The reason for the exchange of $d_{xz}$, $d_{yz}$ and $d_{x^2-y^2}$ orbitals positions in Fe@PtN$_2$ and Fe@PtN$_2$-Cl might be attributed to the fact that after the adsorption of Cl atom, the Fe atom is dragged further away from the PtN$_2$ plane, which can be demonstrated by the N-TM-N bond angle (Table S1). For Fe@PtN$_2$ and Fe@PtN$_2$-Cl systems, although they all have the unpaired electron in Fe-$3d_{z^2}$ orbital, the latter has three unpaired electrons, in line with a lager magnetic moment of 2.65 $\mu_B$ (Fig. 4(a)), thus decreasing the Fe-H interaction (middle penal of Fig. S11). For the Ru@PtN$_2$ system, the unpaired electron of Ru-$4d_{z^2}$ orbital disappears after the adsorption of Cl atom, which will weaken the adsorption of H atom, in line with the conclusion mentioned above (Fig. S10, S11). While for the case of Os@PtN$_2$, an unpaired electron will be induced by the adsorption of Cl atom. However, the adsorption of H atom is not enhanced in Os@PtN$_2$-Cl. This is because the interaction of H atom and the unpaired electron is very weak. Unlike the magnetic systems mentioned above, the unpaired electron in Os@PtN$_2$-Cl is not in the $d_{z^2}$ orbital but in the $d_{x^2-y^2}$ orbital, as depicted in Fig. 4(b). This can be further verified by the PDOS of Os atom that the spin state near Fermi level is mainly contributed by $d_{x^2-y^2}$ orbital (Fig. 4(c)). According to the principle of maximum overlap in molecular orbital theory, the 1s orbital of the H atom attaches to the top site of Os atom is more likely to bond with Os-5$d_{z^2}$ orbital, although the unpaired electron exists in $d_{x^2-y^2}$ orbital, which can be proved by the PDOS after adsorbing H atom (Fig. 4 (d)). Meantime, the magnetic moment of the Os@PtN$_2$-Cl-H* atom does not become zero, illuminating that the unpaired electron is not involved in the interaction with H 1s orbital. That is to say, the spin state contributed by Os-5$d_{x^2-y^2}$ orbital is the "inert spin state", which has little effect on the adsorbed H atom.

The Sabatier principle states that the interaction between the active site and the reactive species should be neither too strong nor too weak in a catalytic reaction [53,54]. If the interaction is too weak, the reaction species is difficult to combine with the catalyst. In view of the weak adsorption of H atoms in non-magnetic systems, the unpaired electron along the out-of-plane direction in magnetic systems acts as an active spin state that can increase the binding energy of H atom. At some cases, the adsorption strength of H atom would reach an intermediate binding energy to permit a compromise between these extremes, thus improving the catalytic performance. A typical case is Pd@PtN$_2$-Cl system with excellent HER activity of a $\Delta G_H$ value of 0.08 eV, even better than the benchmark Pt.

To gain a deeper insight into the strengthening mechanism of H atom adsorption by the active spin state, molecular orbital theory is used to investigate their interactions. For non-magnetic systems (Ni@PtN$_2$, Pd@PtN$_2$, and PtN$_2$), their $d_{z^2}$ orbitals contain two electrons, which will form one bonding orbital σ and one antibonding orbital σ* when they interact with the H-1s orbital. In this case, the antibonding orbital σ* will be occupied (left panel of Fig. 5(a)). On the contrary, for magnetic systems (Co@PtN$_2$, Rh@PtN$_2$, Ir@PtN$_2$, Ni@PtN$_2$-Cl, Pd@PtN$_2$-Cl, and PtN$_2$-Cl), their $d_{z^2}$ orbital only contains one electron, and when the unpaired electron interacts with the H-1s orbital, the formed bonding orbital σ is just enough to accommodate two electrons, leaving no electrons to fill the antibonding orbital (right panel of Fig. 5(a)). According to molecular orbital theory, the fewer antibonding electrons, the more stable the molecule [55]. For both the magnetic and non-magnetic systems, their bonding orbitals are fully occupied; but the magnetic systems have fewer electrons in their antibonding orbital, meaning the larger binding energy of H atom, in agreement with our calculation results of $\Delta G_H$. Taking the non-magnetic Ni@PtN$_2$ and magnetic Co@PtN$_2$, Ni@PtN$_2$-Cl as examples, the negative crystal orbital Hamilton populations (-COHP) of TM-H bond are calculated to verify the theory, as shown in Fig. 5 (b). It can be seen that after inducing an active spin state, the position of the antibonding orbital moves up significantly, demonstrating that the number of electrons occupied on the antibonding orbital is reduced, which is consistent with the principle mentioned above. The -COHP of other systems (Pd@PtN$_2$, Rh@PtN$_2$, Pd@PtN$_2$-Cl, PtN$_2$, Ir@PtN$_2$, and PtN$_2$-Cl) are also confirmed this (Fig. S12).

## Ⅳ. Conclusions

By means of DFT calculations, we have studied the spin-activity correlation based on a series of PtN$_2$-based MSACs with definite magnetic active sites. Although the planar quadrilateral PtN$_2$ monolayer is a non-magnetic material, magnetic active sites can be created by doping transition metal (TM = Fe, Co, Ni, Ru, Rh, Pd, Os, or Ir) atom to replace a Pt atom, or further adsorbing a Cl atom. We reveal that the unpaired electron along the out-of-plane direction such as $d_{z^2}$ orbital of TM atom is an active spin state that will bond with H atom to strengthen the H binding energy because $d_{z^2}$ orbital is just enough to accommodate two electrons to form a bonding orbital. On the contrary, the in-plane unpaired electron such as in $d_{x^2-y^2}$ orbital is inert spin state that has very weak influence on the adsorbed hydrogen atom. This discovery is expected to provide a reference for the design of high-performance catalysts through spin-engineering.

## Acknowledge

This work was supported by the National Natural Science Foundation of China (Grants No. 52172088) and Natural Science Foundation of Hunan Province (No. 2021JJ30112).

## References

[1] B. Yoon, H. Häkkinen, U. Landman, A. S. Wörz, J.-M. Antonietti, S. Abbet, K. Judai, and U. Heiz, Charging Effects on Bonding and Catalyzed Oxidation of CO on $Au_8$ Clusters on MgO, Science **307**, 403 (2005).

[2] N. Mulakaluri, R. Pentcheva, M. Wieland, W. Moritz, and M. Scheffler, Partial Dissociation of Water on $Fe_3O_4$ (001): Adsorbate Induced Charge and Orbital Order, Phys. Rev. Lett. **103**, 176102 (2009).

[3] T. J. Mills, F. Lin, and S. W. Boettcher, Theory and Simulations of Electrocatalyst-Coated Semiconductor Electrodes for Solar Water Splitting, Phys. Rev. Lett. **112**, 148304 (2014).

[4] A. Secchi and F. Troiani, Theory of Multidimensional Quantum Capacitance and Its Application to Spin and Charge Discrimination in Quantum Dot Arrays, Phys. Rev. B **107**, 155411 (2023).

[5] Z. Wang, Y. X. Jiang, J. X. Yin, Y. Li, G. Y. Wang, H. L. Huang et al., Electronic Nature of Chiral Charge Order in the Kagome Superconductor $CsV_3Sb_5$, Phys. Rev. B **104**, 075148 (2021).

[6] H. Oshima and Y. Fuji, Charge Fluctuation and Charge-Resolved Entanglement in a Monitored Quantum Circuit with U (1) Symmetry, Phys. Rev. B **107**, 014308 (2023).

[7] J.-Y. Wang, G.-Y. Huang, S. Huang, J. Xue, D. Pan, J. Zhao, and H. Xu, Anisotropic Pauli Spin-Blockade Effect and Spin–Orbit Interaction Field in an InAs Nanowire Double Quantum Dot, Nano Lett. **18**, 4741 (2018).

[8] X. Liu, K. Wang, T. Zhang, H. Liu, A. Ren, S. Ren, P. Li, C. Zhang, J. Yao, and Y. S. Zhao, Exciton Chirality Transfer Empowers Self-Triggered Spin-Polarized Amplified Spontaneous Emission from 1D-anchoring-3D Perovskites, Advanced Materials 2305260 (2023).

[9] A. Droghetti, M. M. Radonjić, L. Chioncel, and I. Rungger, Dynamical Mean-Field Theory for Spin-Dependent Electron Transport in Spin-Valve Devices, Phys. Rev. B **106**, 075156 (2022).

[10] K. Takeda, A. Noiri, T. Nakajima, T. Kobayashi, and S. Tarucha, Quantum Error Correction with Silicon Spin Qubits, Nature **608**, 682 (2022).

[11] W. Han, S. Maekawa, and X.-C. Xie, Spin Current as a Probe of Quantum Materials, Nat. Mater. **19**, 2 (2020).

[12] T. Bernat, J. S. Meyer, and M. Houzet, Spin Susceptibility of Nonunitary Spin-Triplet Superconductors, Phys. Rev. B **107**, 134520 (2023).

[13] Z. Yang, P. A. Crowell, and V. S. Pribiag, Spin-Helical Detection in a Semiconductor Quantum Device with Ferromagnetic Contacts, Phys. Rev. B **106**, 115414 (2022).

[14] W. Xiong, M. Tian, G.-Q. Zhang, and J. Q. You, Strong Long-Range Spin-Spin Coupling via a Kerr Magnon Interface, Phys. Rev. B **105**, 245310 (2022).

[15] K. Yang et al., Tunable Giant Magnetoresistance in a Single-Molecule Junction, Nat Commun **10**, 3599 (2019).

[16] J. Liang, W. Wang, H. Du, A. Hallal, K. Garcia, M. Chshiev, A. Fert, and H. Yang, Very Large Dzyaloshinskii-Moriya Interaction in Two-Dimensional Janus Manganese Dichalcogenides and Its Application to Realize Skyrmion States, Phys. Rev. B 101, 184401 (2020).

[17] W. Xiong, J. Chen, B. Fang, M. Wang, L. Ye, and J. Q. You, Strong Tunable Spin-Spin Interaction in a Weakly Coupled Nitrogen Vacancy Spin-Cavity Electromechanical System, Phys. Rev. B 103, 174106 (2021).

[18] H. Bai et al., Advances in Spin Catalysts for Oxygen Evolution and Reduction Reactions, Small 19, 2205638 (2023).

[19] D. Xue, P. Yuan, S. Jiang, Y. Wei, Y. Zhou, C.-L. Dong, W. Yan, S. Mu, and J.-N. Zhang, Altering the Spin State of Fe-N-C through Ligand Field Modulation of Single-Atom Sites Boosts the Oxygen Reduction Reaction, Nano Energy 105, 108020 (2023).

[20] Z. Sun et al., Regulating the Spin State of $Fe^{III}$ Enhances the Magnetic Effect of the Molecular Catalysis Mechanism, J. Am. Chem. Soc. 144, 8204 (2022).

[21] L. Li et al., Spin-Polarization Strategy for Enhanced Acidic Oxygen Evolution Activity, Advanced Materials 2302966 (2023).

[22] P. Lv, W. Lv, D. Wu, G. Tang, X. Yan, Z. Lu, and D. Ma, Ultrahigh-Density Double-Atom Catalyst with Spin Moment as an Activity Descriptor for the Oxygen-Reduction Reaction, Phys. Rev. Applied 19, 054094 (2023).

[23] T. Wu et al., Spin Pinning Effect to Reconstructed Oxyhydroxide Layer on Ferromagnetic Oxides for Enhanced Water Oxidation, Nat Commun 12, 3634 (2021).

[24] F. A. Garcés-Pineda, M. Blasco-Ahicart, D. Nieto-Castro, N. López, and J. R. Galán-Mascarós, Direct Magnetic Enhancement of Electrocatalytic Water Oxidation in Alkaline Media, Nat Energy 4, 519 (2019).

[25] X. Ren et al., Spin-Polarized Oxygen Evolution Reaction under Magnetic Field, Nat Commun 12, 2608 (2021).

[26] D. Wu et al., Spin Manipulation in a Metal-Organic Layer through Mechanical Exfoliation for Highly Selective $CO_2$ Photoreduction, Angew Chem Int Ed 62, e202301925 (2023).

[27] J. Yan, Y. Wang, Y. Zhang, S. Xia, J. Yu, and B. Ding, Direct Magnetic Reinforcement of Electrocatalytic ORR/OER with Electromagnetic Induction of Magnetic Catalysts, Adv. Mater. 33, 2007525 (2021).

[28] Z. Li et al., Tuning the Spin Density of Cobalt Single-Atom Catalysts for Efficient Oxygen Evolution, ACS Nano 15, 7105 (2021).

[29] J. Wan et al., Defect Effects on $TiO_2$ Nanosheets: Stabilizing Single Atomic Site Au and Promoting

Catalytic Properties, Advanced Materials 30, 1705369 (2018).

[30] B. Jiang et al., Dynamically Confined Single-Atom Catalytic Sites within a Porous Heterobilayer for CO Oxidation via Electronic Antenna Effects, Phys. Rev. B 107, 205421 (2023).

[31] Q. Fu and C. Draxl, Hybrid Organic-Inorganic Perovskites as Promising Substrates for Pt Single-Atom Catalysts, Phys. Rev. Lett. 122, 046101 (2019).

[32] K. Ding, A. Gulec, A. M. Johnson, N. M. Schweitzer, G. D. Stucky, L. D. Marks, and P. C. Stair, Identification of Active Sites in CO Oxidation and Water-Gas Shift over Supported Pt Catalysts, Science 350, 189 (2015).

[33] Z. Fu, B. Yang, and R. Wu, Understanding the Activity of Single-Atom Catalysis from Frontier Orbitals, Phys. Rev. Lett. 125, 156001 (2020).

[34] Y. Wang, X. Ren, B. Jiang, M. Deng, X. Zhao, R. Pang, and S. F. Li, Synergetic Catalysis of Magnetic Single-Atom Catalysts Confined in Graphitic-$C_3N_4$/$CeO_2$ (111) Heterojunction for CO Oxidization, J. Phys. Chem. Lett. 13, 6367 (2022).

[35] L. Zhang, X. Ren, X. Zhao, Y. Zhu, R. Pang, P. Cui, Y. Jia, S. Li, and Z. Zhang, Synergetic Charge Transfer and Spin Selection in CO Oxidation at Neighboring Magnetic Single-Atom Catalyst Sites, Nano Lett. 22, 3744 (2022).

[36] T. Sun et al., Ferromagnetic Single-Atom Spin Catalyst for Boosting Water Splitting, Nat. Nanotechnol. 18, 763 (2023).

[37] R. Kronberg and K. Laasonen, Reconciling the Experimental and Computational Hydrogen Evolution Activities of Pt (111) through DFT-Based Constrained MD Simulations, ACS Catal. 11, 8062 (2021).

[38] W. Kohn and L. J. Sham, Self-Consistent Equations Including Exchange and Correlation Effects, Phys. Rev. 140, A1133 (1965).

[39] G. Kresse and J. Hafner, Ab Initio Molecular-Dynamics Simulation of the Liquid-Metal–Amorphous-Semiconductor Transition in Germanium, Phys. Rev. B 49, 14251 (1994).

[40] G. Kresse and J. Furthmüller, Efficient Iterative Schemes for Ab Initio Total-Energy Calculations Using a Plane-Wave Basis Set, Phys. Rev. B 54, 11169 (1996).

[41] P. E. Blöchl, Projector Augmented-Wave Method, Phys. Rev. B 50, 17953 (1994).

[42] J. P. Perdew, K. Burke, and M. Ernzerhof, Generalized Gradient Approximation Made Simple, Phys. Rev. Lett. 77, 3865 (1996).

[43] S. Grimme, J. Antony, S. Ehrlich, and H. Krieg, A Consistent and Accurate Ab Initio Parametrization of Density Functional Dispersion Correction (DFT-D) for the 94 Elements H-Pu, The Journal of Chemical Physics 132, 154104 (2010).

[44] S. Grimme, S. Ehrlich, and L. Goerigk, Effect of the Damping Function in Dispersion Corrected Density Functional Theory, J Comput Chem 32, 1456 (2011).

[45] H. J. Monkhorst and J. D. Pack, Special Points for Brillouin-Zone Integrations, Phys. Rev. B 13, 5188 (1976).

[46] G. J. Martyna, M. L. Klein, and M. Tuckerman, Nosé–Hoover Chains: The Canonical Ensemble via Continuous Dynamics, The Journal of Chemical Physics 97, 2635 (1992).

[47] J. Heyd, J. E. Peralta, G. E. Scuseria, and R. L. Martin, Energy Band Gaps and Lattice Parameters Evaluated with the Heyd-Scuseria-Ernzerhof Screened Hybrid Functional, The Journal of Chemical Physics 123, 174101 (2005).

[48] J. Heyd, G. E. Scuseria, and M. Ernzerhof, Hybrid Functionals Based on a Screened Coulomb Potential, The Journal of Chemical Physics 118, 8207 (2003).

[49] V. Wang, N. Xu, J.-C. Liu, G. Tang, and W.-T. Geng, VASPKIT: A User-Friendly Interface Facilitating High-Throughput Computing and Analysis Using VASP Code, Computer Physics Communications 267, 108033 (2021).

[50] V. L. Deringer, A. L. Tchougréeff, and R. Dronskowski, Crystal Orbital Hamilton Population (COHP) Analysis as Projected from Plane-Wave Basis Sets, J. Phys. Chem. A 115, 5461 (2011).

[51] R. Nelson, C. Ertural, J. George, V. L. Deringer, G. Hautier, and R. Dronskowski, LOBSTER: Local Orbital Projections, Atomic Charges, and Chemical-Bonding Analysis from Projector-Augmented-Wave-Based Density□Functional Theory, J Comput Chem 41, 1931 (2020).

[52] See Supplemental Material for the calculated phonon spectrum and AIMD simulations of $PtN_2$; the possible sites for atomic hydrogen adsorption on $PtN_2$; the formation energies of different TM@$PtN_2$; the projected DOS of (TM@)$PtN_2$(-Cl)(-H*); the electron configuration of TM atom in TM@$PtN_2$(-Cl); Bader charges, the negative COHP and the N-TM-N bond angles of (TM@)$PtN_2$-Cl.

[53] A. B. Laursen, A. S. Varela, F. Dionigi, H. Fanchiu, C. Miller, O. L. Trinhammer, J. Rossmeisl, and S. Dahl, Electrochemical Hydrogen Evolution: Sabatier's Principle and the Volcano Plot, J. Chem. Educ. 89, 1595 (2012).

[54] H. Ooka, J. Huang, and K. S. Exner, The Sabatier Principle in Electrocatalysis: Basics, Limitations, and Extensions, Front. Energy Res. 9, 654460 (2021).

[55] Hossain, M.D.; Liu, Z.; Zhuang, M.; Yan, X.; Xu, G.-L.; Gadre, C.A.; Tyagi, A.; Abidi, I.H.; Sun, C.-J.; Wong, H.; et al. Rational Design of Graphene-Supported Single Atom Catalysts for Hydrogen Evolution Reaction. Adv. Energy Mater. 9, 1803689 (2019).

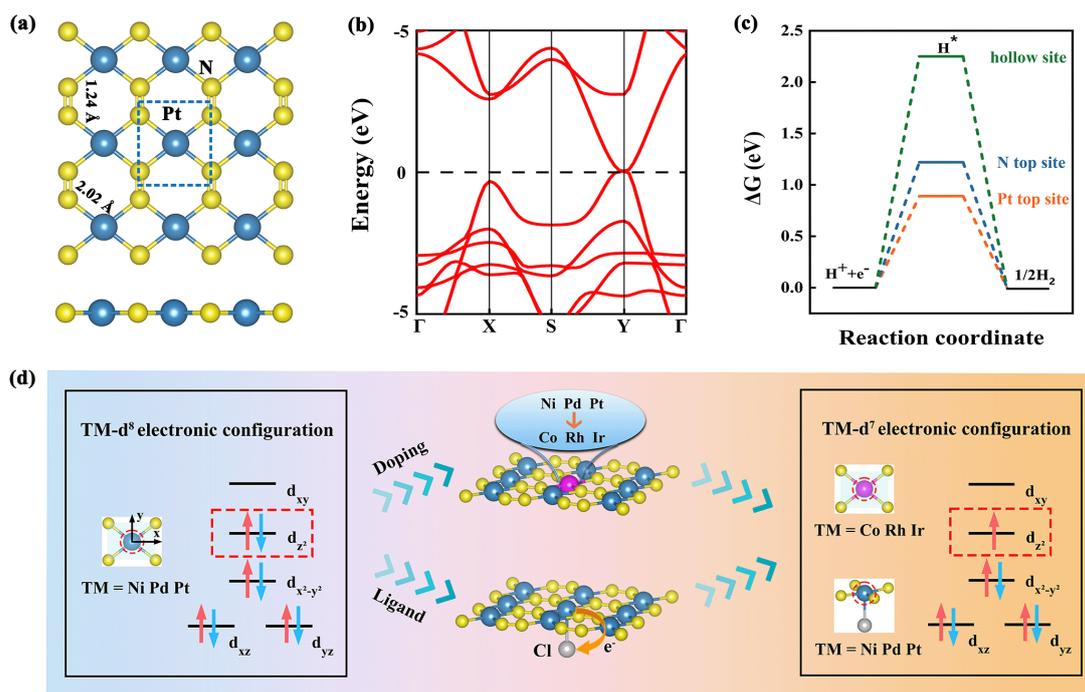

**FIG. 1.** (a) Top and side views of PtN$_2$ monolayer with primitive unit cells shown by the dashed blue line. (b) Electronic band structures of PtN$_2$. The Fermi level is set to 0 eV and shown as dashed black lines. (c) Calculated $\Delta G_{H^*}$ for hydrogen adsorption on different sites of the PtN$_2$ monolayer. (d) The electron configurations of the central atom with and without magnetism induced by two schemes.

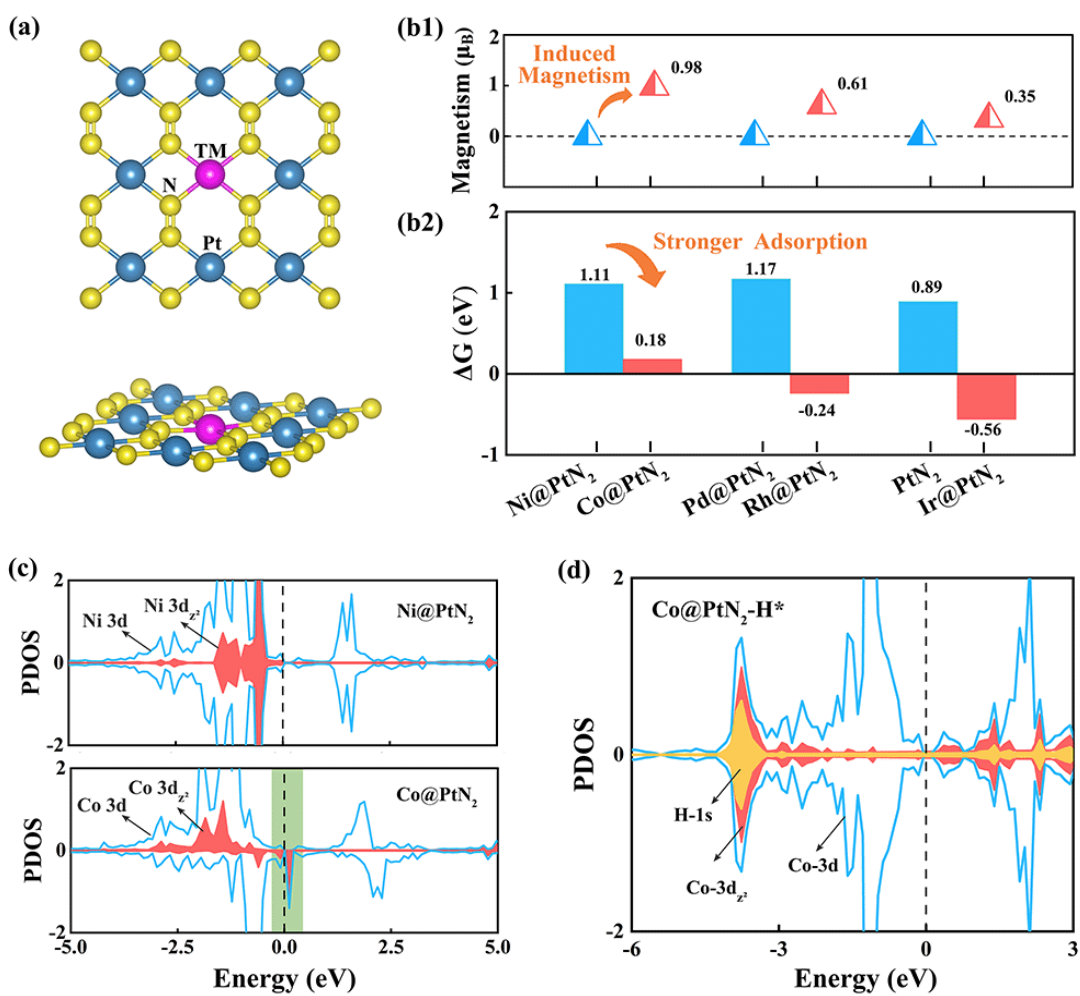

**FIG. 2.** (a) Top and side views of TM@PtN$_2$ monolayer (TM = Ni, Co, Pd, Rh, Pt, Ir). (b1) Magnetic moments of TM@PtN$_2$ with the dashed black line representing the magnetic moment of zero. (b2) Calculated $\Delta G_H$ on TM@PtN$_2$. (c) The PDOS of Ni@PtN$_2$ and Co@PtN$_2$. (d) The PDOS of Co@PtN$_2$ after H adsorbed on Co top site. The Fermi levels are set to 0 eV and shown as dashed black lines.

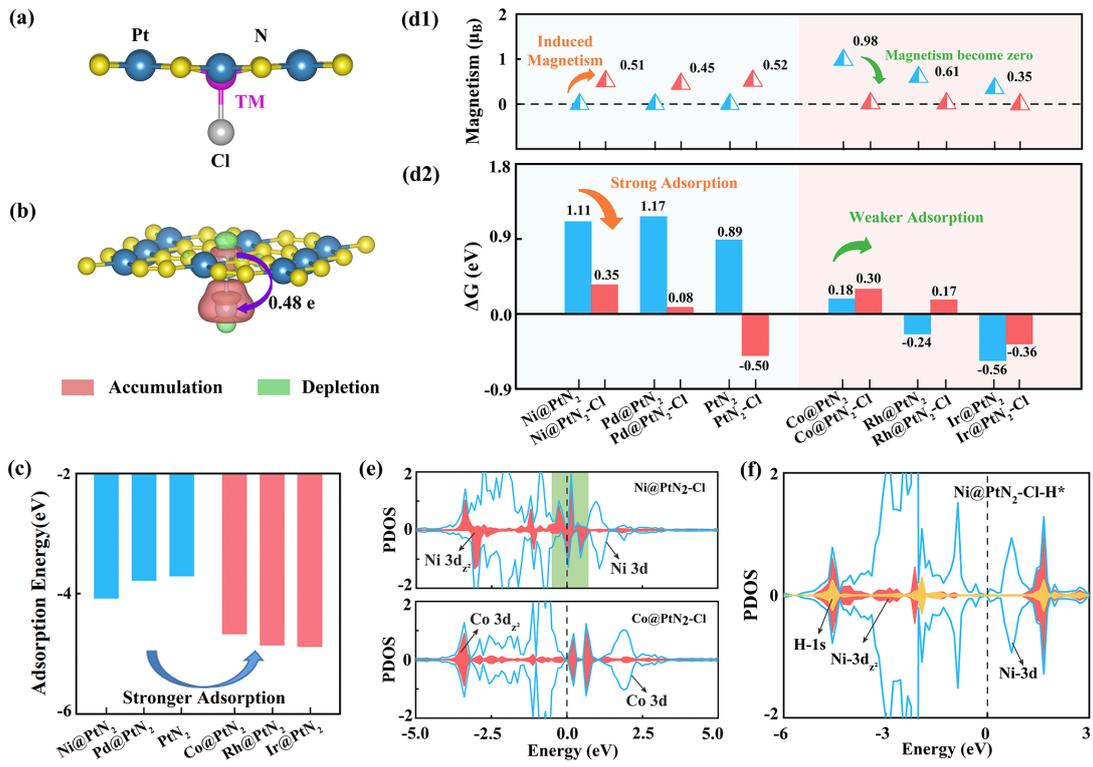

**FIG. 3.** (a) Side view of TM@PtN$_2$-Cl monolayer (TM = Ni, Pd, Pt, Co, Rh, Ir). (b) The charge distribution between Ni and Cl of Ni@PtN$_2$-Cl. (c) Adsorption energy of Cl atom in different TM@PtN$_2$. (d1) Magnetic moments of TM@PtN$_2$ and TM@PtN$_2$-Cl with the dashed black line representing the magnetic moment of zero. (d2) Calculated $\Delta G_H$ on TM@PtN$_2$ and TM@PtN$_2$-Cl. (e) The PDOS of Ni@PtN$_2$-Cl and Co@PtN$_2$-Cl. (f) The PDOS of Ni@PtN$_2$-Cl after H adsorbed on Ni top site. The Fermi levels are set to 0 eV and shown as dashed black lines.

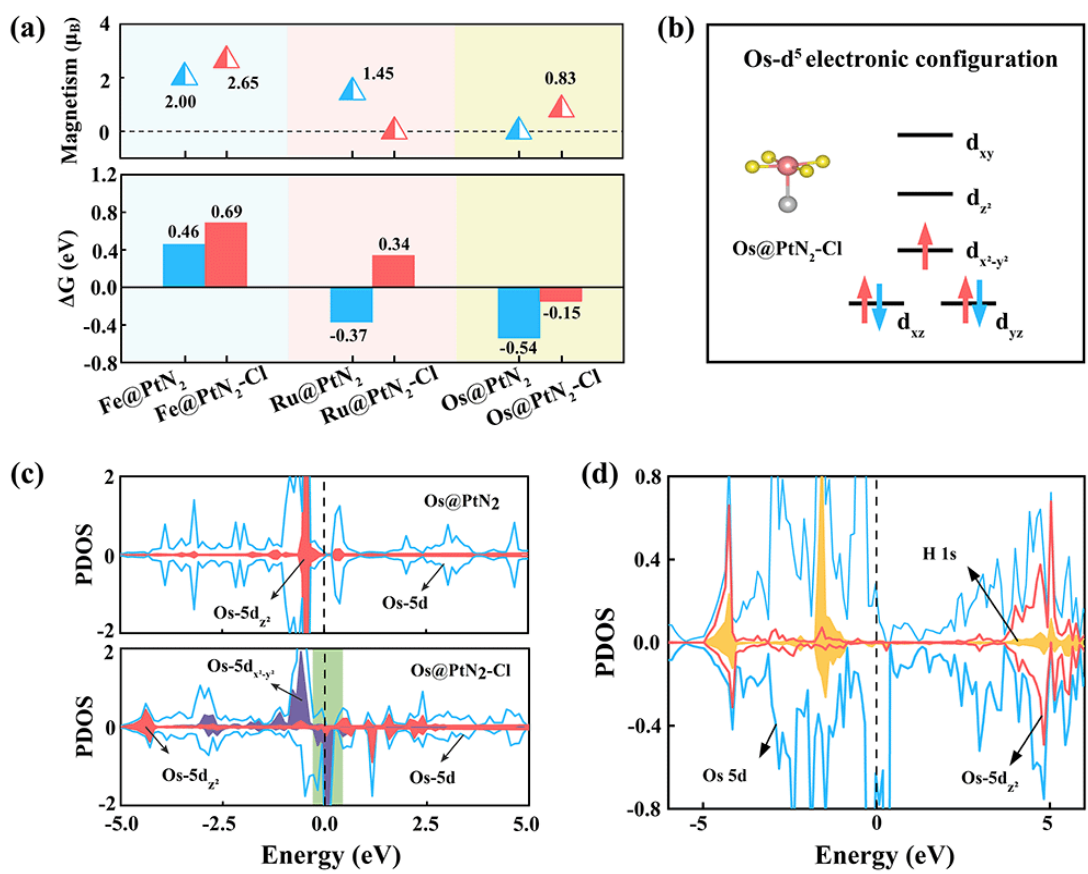

**FIG. 4.** (a) The magnetic moment and $\Delta G_H$ of TM@PtN$_2$ and TM@PtN$_2$-Cl (TM = Fe, Ru, Os) with the dashed black line representing the magnetic moment of zero. (b) Os 5d orbital electronic configuration in Os@PtN$_2$-Cl. (c) The PDOS of Os@PtN$_2$ and Os@PtN$_2$-Cl. (d) The PDOS of Os@PtN$_2$-Cl after H adsorbed on Os top site. The Fermi levels are set to 0 eV and shown as dashed black lines.

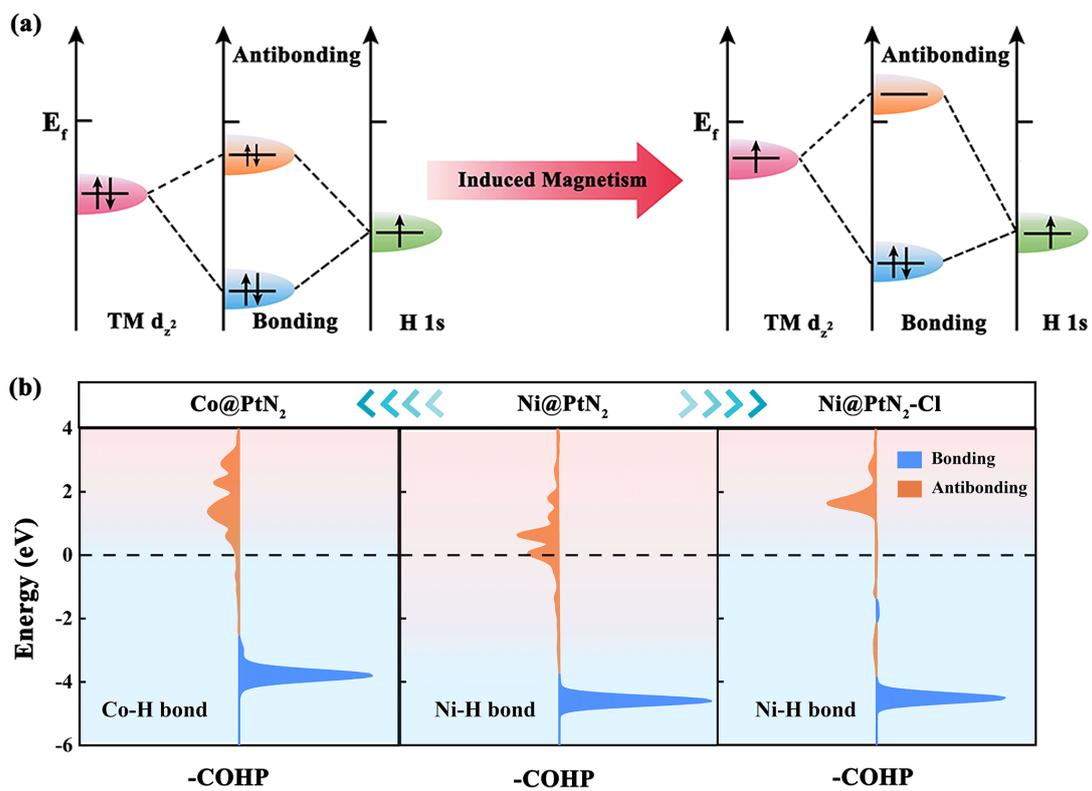

**FIG. 5.** (a) The combination of H-1s orbital and TM-$dz^2$ orbital forming a bonding orbital and an anti-bonding orbital, in which the occupancy of anti-bonding orbital will determine the strength of H–TM bond. (b) The negative crystal orbital Hamilton populations (-COHP) of TM-H bond in Ni@PtN$_2$, Co@PtN$_2$, Ni@PtN$_2$-Cl, respectively. The Fermi levels are set to 0 eV and shown as dashed black lines.